\documentclass[showpacs,prd]{revtex4}
\usepackage{epsfig}

\begin{document}

\newcommand{\be}{\begin{enumerate}}
\newcommand{\ee}{\end{enumerate}}

\title{\large \boldmath Evidence for $f_0(980)f_0(980)$ production  in $\chi_{c0}$ decays}
\author{\small 
M.~Ablikim$^{1}$, J.~Z.~Bai$^{1}$, Y.~Ban$^{10}$,
J.~G.~Bian$^{1}$, X.~Cai$^{1}$, J.~F.~Chang$^{1}$,
H.~F.~Chen$^{16}$, H.~S.~Chen$^{1}$, H.~X.~Chen$^{1}$,
J.~C.~Chen$^{1}$
% ...
\vspace{0.1cm}
\\(BES Collaboration)\\
\vspace{0.2cm}
$^1$ Institute of High Energy Physics, Beijing 100039, People's
Republic of     China\\
% ...
$^{10}$ Peking University, Beijing 100871, People's Republic of
China\\
% ...
$^{16}$ University of Science and Technology of China, Hefei 230026,
        People's Republic of China\\
% ...
}

\vspace*{0.4cm}
\date{\today}

\begin{abstract}
Using a sample of 14 million $\psi(2S)$ events accumulated with the
BES\,II detector, evidence for $f_0(980)f_0(980)$ production in
$\chi_{c0}$ decays is obtained for the first time; the branching ratio
is determined to be ${\cal B}(\chi_{c0}\to
f_0(980)f_0(980)\to\pi^+\pi^-\pi^+\pi^-) = (7.6 \pm 1.9~({\rm stat}) \pm
1.6~ ({\rm syst})) \times 10^{-4}$. The significance of the $f_0(980)$
signal is about 4.6$\sigma$.
\end{abstract}

\maketitle


\section{Introduction}

After thirty years of controversy, the nature of the $f_0(980)$ is
still not settled~\cite{REVIEW}.  It has been described as a
conventional $q\bar q$ meson \cite{Morgan}, a ``unitarized remnant''
of a $q\bar q$ state \cite{Tor}, a $K\bar K$ molecule \cite{Wein}, a
multiquark state \cite{Jaffe}, or a glueball \cite{Jaffe2}.  Because
of its close proximity to the $K\bar K$ threshold and its propensity
to decay to $K\bar K$, it is difficult to quantify even the mass and
width of the $f_0(980)$. To be explicit, the state with a mass of $980 \pm
10$ MeV and a width somewhere between 40 and 100 MeV \cite{PDG}
straddles the $K\bar K$ threshold at 990 MeV.  Many arguments favoring
or disfavoring the above assignments depend on the width or pole
position of the $f_0(980)$.

A novel measurement to elucidate the
nature of the $f_0(980)$ was suggested by Refs. [8,9]. By determining
the radiative decay rate for $\phi\to f_0(980)\gamma$, one can infer
the $s\bar s$ content of the $f_0$ wave function since the rate is
proportional to the overlap with the $\phi$, a well known $s\bar s$
state.  The results from CMD2, SND, and KLOE \cite{980} reveal a much
higher branching ratio for radiative $\phi\to\gamma f_0$ decay than
that expected for the $q\bar q$ meson or $K\bar K$ molecule
interpretations. It seems that these data add weight to the idea that
the $f_0(980)$ is a compact $qq\bar q\bar q$ state with an extended
meson-meson cloud `molecular' tail \cite{FEC2}. However, at present
the interpretation about the nature of the $f_0(980)$ is still open
\cite{PDG}, and more experimental results are needed to clarify it.

In this paper, we report on the analysis of $\pi^+\pi^-\pi^+\pi^-$
final states from $\chi_{c0}$ decays using a sample of 14 million
$\psi(2S)$ events accumulated with the BES\,II detector.  Evidence for
$f_0(980)f_0(980)$ production from $\chi_{c0}$ decays is obtained for
the first time.

\section{Bes detector}
BES\,II is a large
solid-angle magnetic spectrometer that is described in detail in Ref.
\cite{BESII}. Charged particle momenta are determined with a
resolution of $\sigma_p/p = 1.78\%\sqrt{1+p^2}$~($p$~in~GeV$/c$) in a
40-layer cylindrical drift chamber (MDC). Particle identification is
accomplished by specific ionization ($dE/dx$) measurements in the
drift chamber and time-of-flight (TOF) measurements in a barrel-like
array of 48 scintillation counters. The $dE/dx$ resolution is
$\sigma_{dE/dx} = 8.0\%$; the TOF resolution is $\sigma_{TOF} = 180$
ps for Bhabha events. Outside of the time-of-flight counters is a
12-radiation-length barrel shower counter (BSC) comprised of gas
proportional tubes interleaved with lead sheets. The BSC measures the
energies of photons with a resolution of
$\sigma_E/E\simeq 21\%/\sqrt{E}$~($E$ in GeV). Outside the solenoidal
coil, which provides a 0.4 Tesla magnetic field over the tracking
volume, is an iron flux return that is instrumented with three double layers of counters that are
used to identify muons.

In this analysis, a
 GEANT3 based Monte Carlo simulation package (SIMBES) with detailed
   consideration of detector performance (such as dead
   electronic channels) is used.
    The consistency between data and Monte Carlo has been
 checked in
   many high purity physics channels, and the agreement is quite
 reasonable.


\section{Event selection}

The selection criteria used in this analysis are similar to those
of Ref. \cite{BESc}; the main difference between them is that no
particle identification is imposed here in order to increase the
selection efficiency.

\subsection{Photon Identification}

A neutral cluster is considered to be a photon candidate when the
angle between the nearest charged track and the cluster is greater
than 15$^{\circ}$, the first hit is in the beginning 6 radiation
lengths, and the difference between the angle of the cluster
development direction in the BSC and the photon emission direction is
less than 30$^{\circ}$. The photon candidate with the largest energy
deposit in the BSC is treated as the photon radiated from the $\psi(2S)$
and used in a four-constraint kinematic fit to the hypothesis
$\psi(2S)\to\gamma\pi^+\pi^-\pi^+\pi^-$.

\subsection{Charged Particle Identification}

Each charged track is required to be well fit to a three-dimensional
helix using the MDC information, be in the polar angle region
$|\cos\theta_{{MDC}}| < 0.80$, and have the point of closest approach
of the track to the beam axis be within 2 cm of the beam
and within 20 cm of the center of the interaction region along the
beam line.

\subsection{Event Selection Criteria}

The candidate events are
required to satisfy the following selection criteria:

\be
\item  The number of charged tracks is required to be four with net
charge zero.

\item The sum of the momenta of the lowest momentum $\pi^+$ and
$\pi^-$ tracks is required to be greater than 650 MeV; this removes
contamination from $\psi(2S)\to\pi^+\pi^- J/\psi$ and some 
          $\rho^0\pi\pi$ events.

\item The $\chi^2$ probability for the four-constraint kinematic fit to
the decay hypothesis $\psi(2S)\to\gamma\pi^+\pi^-\pi^+\pi^-$ is
greater than 0.01.
\ee

The invariant mass distribution for the $\pi^+\pi^-\pi^+\pi^-$ events
that survive all the selection requirements is shown in Fig. 1. There
are clear peaks corresponding to the $\chi_{cJ}$ states. The highest
mass peak
corresponds to charged track final states that are kinematically
fit with an unassociated photon.

\begin{figure}[hbtp]
\begin{center}
\epsfxsize=6.75cm\epsffile{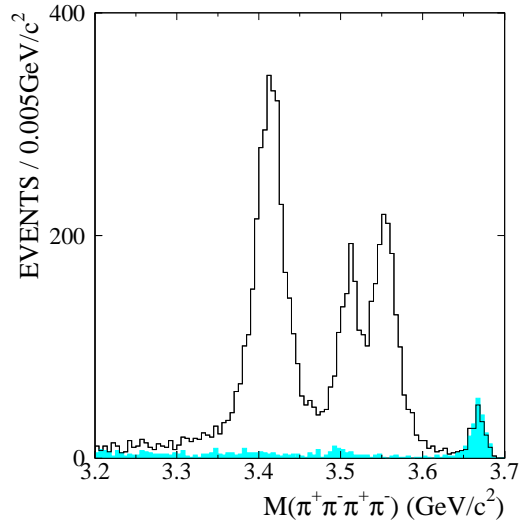}
\label{cpkp}
\caption{The $\pi^+\pi^-\pi^+\pi^-$ invariant mass spectrum. The
shadow histogram shows the spectrum for Monte Carlo simulated background events.}
\end{center}
\end{figure}

The distribution of background events in the $4\pi$ mass spectrum,
determined by Monte Carlo simulation normalized using PDG2004 branching
ratios \cite{PDG}, is also shown in Fig. 1. The distribution is flat in
the mass range of the $\chi_{cJ}$ states, and the background events
come mainly from $\psi(2S)\to\pi^0\pi^+\pi^-\pi^+\pi^-$.  The highest
mass peak is from $\psi(2S)\to\pi^+\pi^-\pi^+\pi^-$ and
$\psi(2S)\to\pi^+\pi^-K^+K^-$ events combined with an unassociated
photon.  The background is very low compared with the strong
$\chi_{cJ}$ peaks, and its effect will not be considered in the
following analysis.

In this analysis, no particle identification is imposed on charged
tracks in order to increase the detection efficiency.  It is not
necessary to distinguish pions from kaons or protons in this channel
because the background is not serious, as shown in Fig. 1, and the
contamination from events with kaons or protons is rejected
effectively by the kinematic fit. The statistics of present data
sample is about 20\% higher than one using particle identification.

\section{ANALYSIS RESULTS}

\subsection{\boldmath $f_0(980)f_0(980)$ signal}

Figure 2 shows scatter plots of $\pi^+\pi^-$ versus $\pi^+\pi^-$
invariant mass \cite{pairing} for events with $\pi^+\pi^-\pi^+\pi^-$
mass between 3.30 and 3.48 GeV and between 3.53 and 3.60 GeV, and the
corresponding projections are shown in Fig. 3 (two entries per
event). The clusters of events in the lower left-hand corners of
Figs. 2(a) and 2(b) indicate the presence of a $K^0_SK^0_S$ signal
under both the $\chi_{c0}$ and $\chi_{c2}$ peaks. A clear
$f_0(980)f_0(980)$ signal can be seen in Fig.~2(a). There are some
hints of $\rho^0\rho^0$ and $f_0(1370)f_0(1370)$~(or
$f_2(1270)f_2(1270)$) signals in Fig.~2(a) and $\rho^0\rho^0$ but no
$f_0(980)f_0(980)$ events in Fig.~2(b).  In this paper, we study the
$f_0(980)f_0(980)$ in $\chi_{c0}$ decays.

\vspace*{5pt}

\begin{figure}[htbp]
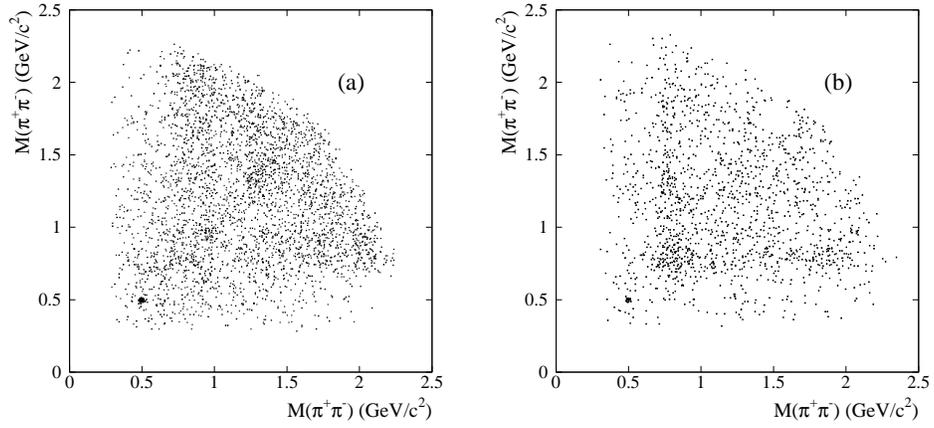

\begin{center}
\epsfxsize=5.7500cm\epsffile{sca-4piic.epsi}
\hspace*{0.5cm}
\epsfxsize=5.7500cm\epsffile{sca-4piid.epsi}
\label{cpkp1}
\caption{Scatter plots of $\pi^+\pi^-$ versus $\pi^+\pi^-$ invariant
mass for selected $\gamma\pi^+\pi^-\pi^+\pi^-$ events with
$\pi^+\pi^-\pi^+\pi^-$ mass in (a) the $\chi_{c0}$  and (b) the $\chi_{c2}$
mass regions.}
\end{center}
\end{figure}

\begin{figure}[htbp]
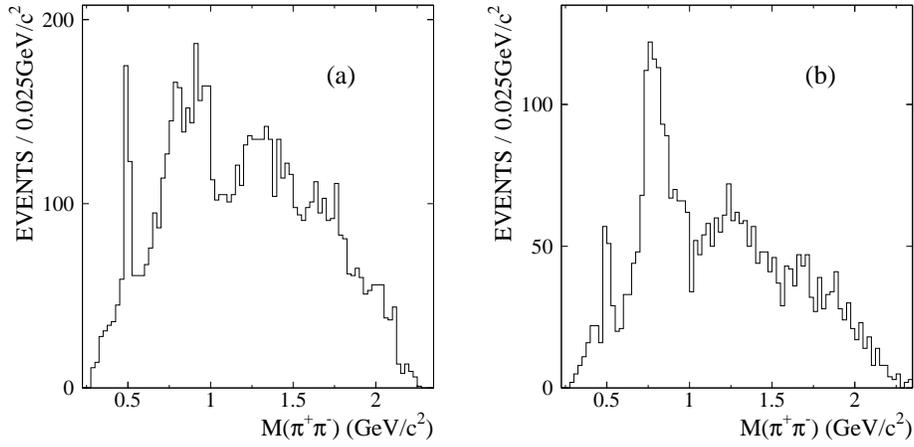

\begin{center}
\epsfxsize=5.6500cm\epsffile{c00.epsi}
\hspace*{0.5cm}
\epsfxsize=5.650cm\epsffile{c22.epsi}
\label{cpkp2}
\caption{Projections of $\pi^+\pi^-$ invariant
mass under the (a) $\chi_{c0}$  and (b) $\chi_{c2}$
peaks (two entries per event).  }
\end{center}
\end{figure}

For the events in $\chi_{c0}$ mass region (from 3.30 to 3.48 GeV) and
after requiring that the mass of one of the $\pi^+\pi^-$ pairs lies
between 0.88 and 1.04 GeV, the mass distribution of the other
$\pi^+\pi^-$ pair is shown in Fig. 4 (two entries per event); there is
a strong $f_0(980)$ signal, and its line shape is similar to other
experiments \cite{PDG}.  From a Monte Carlo simulation, the background
in the $f_0(980)$ region is mainly from processes such as
$\psi(2S)\to\gamma\chi_{c0},~\chi_{c0}\to
a_1(1260)\pi,~a_1(1260)\to\rho\pi$.

\begin{figure}[htbp]
\begin{center}
\epsfxsize=6.5cm\epsffile{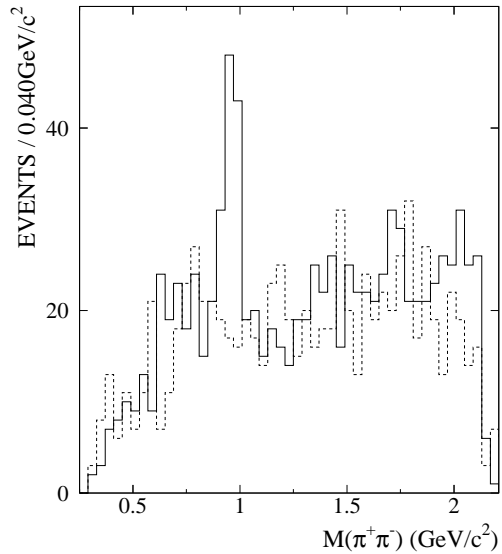}
\label{cpkp3}
\caption{Plot of $\pi^+\pi^-$ mass recoiling against the
$f_0(980)~(0.88$ GeV $ < m_{\pi^+\pi^-} < 1.04 $ GeV) for events in
the $\chi_{c0}$ mass region (two entries per event), where the dashed
line histogram indicates a rough estimation of background determined from
sidebands.}
\end{center}
\end{figure}

\begin{figure}[htbp]
\begin{center}
\epsfxsize=7.0cm\epsffile{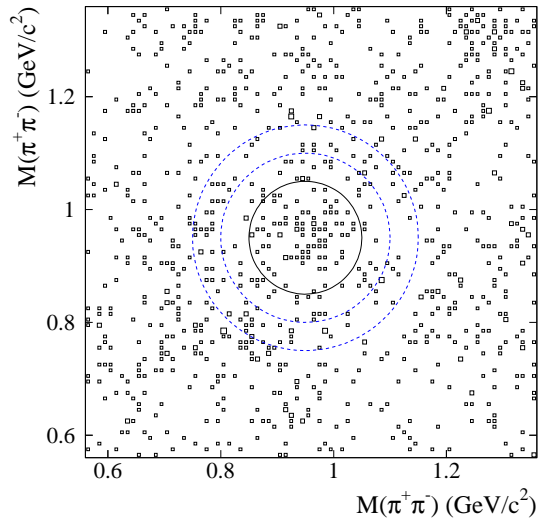}
\label{cpkp4}
\caption{Scatter plot of $\pi^+ \pi^-$ versus $\pi^+ \pi^-$ invariant
  mass in the $f_0(980)$ region for $\chi_{c0}$ candidate events, showing the definition of signal and
  background regions.}
\end{center}
\end{figure}

Note that the background estimation in Fig. 4 using sidebands (0.76 -
0.84 GeV and 1.08 - 1.16 GeV) is rough. In this paper, the number of
$f_0(980)f_0(980)$ events and the corresponding background are
estimated from the scatter plot of $\pi^+\pi^-$ versus $\pi^+\pi^-$
invariant masses, as shown in Fig. 5. This method gives more accurate
determinations of the $f_0(980)$ signal and background.  The signal
region is shown in Fig. 5 as a circle centered at $(0.960, 0.960)$ GeV
and with a radius of 80 MeV, and the background is estimated from the
events between two circles with radii of 120 MeV and 160 MeV.  There
are 65 and 51 events in the signal and background regions,
respectively. So the number of $f_0(980)f_0(980)$ events is estimated
to be 65 - 51/1.75 = $35.9\pm 9.0$, where 1.75 is the normalization
factor -- the ratio of the area of background region to that of the
signal region. The $\pi^+\pi^-$ mass range we adopt is shifted from
the $f_0(980)$ central mass value of 980 MeV because of the asymmetric
character of its mass spectrum. We obtain the signal significance of
the $f_0(980)f_0(980)$ of 4.6$\sigma$ using the method described in
Ref.~\cite{Narsky}.

\subsection{Monte Carlo simulation}

A Monte Carlo simulation is used to determine the detection efficiency. The
angular distribution of the emitted photon in the process
$\psi(2S)\to\gamma\chi_{c0}$ is taken
into account \cite{E1}. The
$f_0(980)$ is generated with the usual Flatt\'e formula \cite {Fla, BS}:
$$ f = {{1}\over{M^2 - s - iM(g_1\rho_1 + g_2\rho_2)}},$$ where
$\rho_{1,2}$ are the phase space factors for the $\pi\pi$ and $K\bar
K$ channels, $\rho_i(s) = \sqrt{1-m^2_i/4s}$, and $g_{1,2}$ are
squares of coupling constants to the two channels. For the $K\bar K$
channel, $m^2$ is taken as the average of the $K^0$ and $K^{\pm}$
masses, and the algebraic expression for $\rho_2$ is extended
analytically below the $K\bar K$ threshold. In the simulation, the
parameters used are those of Ref. \cite{BS}: M = 0.9535, $g_1 =
0.1108$, $g_2 = 0.4229$ GeV.  

\subsection{Branching fraction results}

The efficiency is determined using 100,000 Monte Carlo simulated
events that are passed through the same selection as the data events;
the efficiency is estimated to be $\epsilon = (3.92\pm0.07)\%$, where
the error is the statistical error of the Monte Carlo sample. Note
that for this estimation, the events in the background region are
subtracted from the events in the signal region, similar to the
treatment of data.

Using numbers from
above, the branching ratio of
$\psi(2S)\to\gamma\chi_{c0},~\chi_{c0}\to
f_0(980)f_0(980)\to\pi^+\pi^-\pi^+\pi^-$ is 
$${\cal B}(\psi(2S)\to\gamma\chi_{c0}\to \gamma
f_0(980)f_0(980)\to \gamma\pi^+\pi^-\pi^+\pi^-) = (6.54\pm
1.64)\times 10^{-5},$$
where the error is statistical.

\begin{figure}[htbp]
\begin{center}
\epsfxsize=8.0cm\epsffile{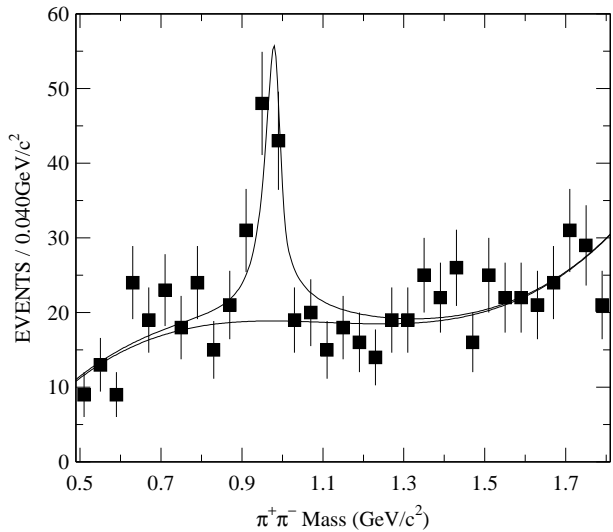}
\label{cpkp5}
\caption{A fit to the $\pi^+\pi^-$ mass recoiling against the
$f_0(980)~(0.88$ GeV $ < m_{\pi^+\pi^-} < 1.04 $ GeV) for events in
the $\chi_{c0}$ mass region (two entries per event).
}
\end{center}
\end{figure}

As a cross-check, a fit to the projected $\pi^+\pi^-$ mass
distribution between 0.49 and 1.81 GeV with the Flatt\'e
parameters of the $f_0(980)$ fixed to the solution of Ref. \cite{BS} plus
a polynomial background yields $84.8\pm 18.4$ events for the
$f_0(980)$ signal, as shown in Fig. 6 ({two entries per event}).
From the results of this fit, we
determine the branching ratio
$${\cal B}(\psi(2S)\to\gamma\chi_{c0}\to
\gamma f_0(980)f_0(980)\to \gamma\pi^+\pi^-\pi^+\pi^-) 
= (4.54\pm
0.98)\times 10^{-5},$$ where the detection efficiency is
13.34\% (two entries per event). Because of  
the low
statistics and the relatively high backgrounds, as well as the lack of
information from the coupled $K\bar K$ channel, a free fit to the
parameters M, $g_1$, and $g_2$ is difficult. The
branching ratio values determined by these two methods agree to about
one sigma.

\subsection{Systematic errors}

The systematic errors in the branching ratio measurement associated
with the efficiency are determined by comparing $\psi(2S)$ data and
Monte Carlo simulation for very clean decay channels, such as
$\psi(2S)\to\pi^+\pi^-J/\psi$, which allows the determination of
systematic errors associated with the MDC tracking, kinematic fitting,
and the photon identification \cite{CZ}.
Other sources of systematic error come from
the uncertainties of the number of $\psi(2S)$ events \cite{moxh}, the
parameters of the $f_0(980)$, the definition of background region, the
$\chi_{c0}$ and $f_0(980)$ mass resolutions, etc. 

\subsubsection{Parameters of the $f_0(980)$}
The parameters of the $f_0(980)$ are still uncertain, and
different descriptions of the $f_0(980)$ in the simulation result
in different  efficiencies. Besides the solution of
Ref. \cite{BS}, we also consider the
measurements
of some recent
experiments such as E791, GAMS and WA102 [21-23], where a 
Breit-Wigner description with the width varying from 44
to 80 MeV was used for the $f_0(980)$.
We determine the change both by using
the solutions of Refs. [21-23] and by varying $g_1$ in
Ref. \cite{BS} from
0.1108 GeV to 0.090 GeV and 0.130 GeV while keeping the ratio $g_2/g_1$
fixed.
The largest change is about 16\%, which is used for the systematic
error due to this uncertainty.

\subsubsection{Different background regions}
In this paper, we estimate the background using the
region between circles with radii
120 MeV and 160 MeV about $(0.960, 0.960)$ GeV,
as shown in Fig. 5. We test two other different background
definitions
by changing the radii of the two circles to 100 and 150 MeV and
120 and 180 MeV. 
The biggest change is about 5\%, which is taken as the systematic error.

\subsubsection{$\chi_{c0}$ and $f_0(980)$ mass resolutions}

Differences between data and Monte Carlo mass resolutions
for the $\chi_{c0}$ and $f_0(980)$ also cause systematic uncertainties
in the determination of the branching ratio of $\chi_{c0}\to
f_0(980)f_0(980)$. From a study, we find that the difference for
the $\chi_{c0}$ is about 1 MeV, so we change the window of $\chi_{c0}$ to
[3.300 + 0.005,~3.480 - 0.005] GeV and [3.300 - 0.005,~3.480 + 0.005]
GeV, and estimate the
effect on the branching ratio.  Such changes result in less than a
1\% variation in the efficiency, and the effect of the difference in
the mass resolutions of the $f_0(980)$ is even smaller. By varying the
width of $\chi_{c0}$ by 1$\sigma$ of its error, 0.8 MeV, there is
almost no change on the detection efficiency.   We include
a 1\% systematic error for the sum of these uncertainties.

Table I lists the systematic errors from all sources,
and adding them in quadrature, the total systematic error, $20\%$, is
obtained. The resulting branching ratio is
$${\cal
B}(\psi(2S)\to\gamma\chi_{c0}\to \gamma f_0(980)f_0(980)\to \gamma\pi^+\pi^-\pi^+\pi^-) = (6.5\pm
1.6\pm 1.3)\times 10^{-5},$$
and  finally
using the PDG2004 average value and error for
${\cal B}(\psi(2S)\to\gamma\chi_{c0})$  \cite{PDG}, we obtain 
$${\cal
B}(\chi_{c0}\to
f_0(980)f_0(980)\to\pi^+\pi^-\pi^+\pi^-) = (7.6\pm
1.9~(\mbox{stat})\pm 1.6~(\mbox{syst}))\times 10^{-4}.$$

\begin{table}[htbp]
\begin{center}
\caption{Summary of systematic errors in the branching ratio
  calculation of ${\cal B}(\psi(2S)\to \gamma \chi_{c0}\to \gamma
f_0(980)f_0(980)\to\gamma\pi^+\pi^-\pi^+\pi^-)$.}
\begin{tabular}{l|r}
\hline
\hline
Source &~~~Relative systematic error\\
\hline
MDC tracking&~~~8\%\\
Kinematic fit&6\%\\
Photon ID efficiency&2\%\\
$\psi(2S)$ number&4\%\\
Efficiency estimation&16\%\\
Definition of background~~~~&5\%\\
Mass
resolutions&1\%\\
\hline
Total &20\%\\
\hline\hline

\end{tabular}
\end{center}
\end{table}

\section{Summary}
Evidence for $f_0(980)f_0(980)$ production from $\chi_{c0}$ decays
is obtained for the first time with a significance of about 4.6$\sigma$
, and
the branching ratio is determined to be ${\cal B}(\chi_{c0}\to
f_0(980)f_0(980)
\to\pi^+\pi^-\pi^+\pi^-) = (7.6 \pm 1.9 ~(\mbox{stat})
\pm 1.6~(\mbox{syst}))\times
10^{-4}$.
This may help in understanding the nature of $f_0(980)$.

\section{Acknowledgments}

   The BES collaboration thanks the staff of the BEPC for their hard
   efforts.
This work is supported in part by the National Natural Science
   Foundation
of China under contracts Nos. 19991480, 10225524, 10225525, the Chinese
   Academy
of Sciences under contract No. KJ 95T-03, the 100 Talents Program of
   CAS
under Contract Nos. U-11, U-24, U-25, and the Knowledge Innovation
   Project of
CAS under Contract Nos. U-602, U-34 (IHEP); by the National Natural
   Science
Foundation of China under Contract No.10175060 (USTC); and by the
   Department
of Energy under Contract No.
DE-FG03-94ER40833 (U Hawaii).


\begin{thebibliography}{99}
\bibitem{pairing} One $\pi^+ \pi^-$ pair is formed using the highest
  momentum $\pi^+$ and the lowest momentum $\pi^-$ tracks.  The other
  pair uses the two remaining tracks.
\end{thebibliography}

\end{document}